\documentclass[11pt,a4paper]{article}
\usepackage{amsmath,amssymb}
\usepackage{epsfig,graphicx}

\begin{document} 

\title{\bf Operation of a GERDA Phase I prototype detector in liquid argon and nitrogen}
\author{M.~Barnab\'e~Heider$^a$\footnote{{\bf e-mail}: Marik.Barnabe-Heider@mpi-hd.mpg.de},
A.~Bakalyarov$^{b}$, L.~Bezrukov$^{c}$, C.~Cattadori$^{d}$, O.~Chkvorets$^{a}$,\\ K.~Gusev$^{b,e}$, M.~Hult$^{f}$, I.~Kirpichnikov$^{g}$, V.~Lebedev$^{b}$, G.~Marissens$^{f}$, P.~Peiffer$^{a}$,\\ S.~Sch\"onert$^{a}$\ M.~Shirchenko$^{b,e}$, A.~Smolnikov$^{c,e}$, A.~Vasenko$^{g}$, S.~Vasiliev$^{c,e}$, S.~Zhukov$^{b}$\\
$^a$ \small{\em Max-Planck-Institut f\"ur Kernphysik, Saupfercheckweg 1, D-69117 Heidelberg, Germany}\\
$^b$ \small{\em Russian Research Center Kurchatov Institute, 123182 Moscow, Russia} \\
$^c$ \small{\em Institute for Nuclear Research, 117312 Moscow, Russia}\\
$^d$ \small{\em Laboratori Nazionali del Gran Sasso, S.S. 17 bis km.18.910, 67010 Assergi (AQ), Italy} \\
$^e$ \small{\em Joint Institute for Nuclear Research, 141980 Dubna, Russia}\\
$^f$ \small{\em EC-JRC-Institute for Reference Materials and Measurements, Retieseweg 111, B-2440 Geel, Belgium} \\
$^g$ \small{\em Institute for Theoretical and Experimental Physics, 117218 Moscow, Russia}
}
\date{April 19, 2007}

\maketitle
\begin{abstract}
The GERDA (GERmanium Detector Array) experiment aiming to search for the neutrinoless double beta decay (0$\nu \beta \beta$) of $^{76}$Ge at the Laboratori Nazionali Del Gran Sasso (LNGS), Italy, will operate bare enriched high-purity germanium (HPGe) detectors in liquid argon. GERDA Phase I will use the enriched diodes from the previous Heidelberg-Moscow (HdM) and IGEX experiments. With the HPGe detectors mounted in a low-mass holder, GERDA aims at an excellent energy resolution and extremely low background. The goal is to check the claim for the 0$\nu \beta \beta$ evidence in the HdM $^{76}$Ge experiment within one year of data taking. Before dismounting the enriched diodes from their cryostat, the performance parameters of the HdM and the IGEX detectors have been measured. The diodes have been removed from their cryostats, their dimensions measured and they have been put under va-cuum in a transportation container. They are now being refurbished for GERDA Phase I at Canberra Semiconductor NV. Before operating the enriched diodes, a non-enriched HPGe p-type detector mounted in a low-mass holder is operated in the liquid argon test facility of the GERDA Detector Laboratory (GDL) at LNGS. Since January 2006, the testing of the prototype detector is being
carried out in the GDL as well as at the
site of the detector manufacturer.
\end{abstract}

\section{Introduction}

The GERDA experiment will search for 0$\nu \beta \beta$ decay of $^{76}Ge$ \cite{gerda}. Bare HPGe detectors enriched in $^{76}$Ge will be submerged in liquid argon serving simultaneously as a shield against external radioactivity and a cooling medium. The first phase of GERDA will operate the existing enriched diodes from the past HdM \cite{hdm} and IGEX \cite{igex} experiments. In total, 8 HPGe detectors (total mass $\sim$18 kg) enriched with $^{76}$Ge at 86 \% and 6 natural germanium diodes from Genius-TF \cite{genius} (total mass $\sim$15 kg) will be operated. GERDA Phase I aims at a total background index of less than $10^{-2}$ cts/(keV$\cdot$·kg$\cdot$·y) in the Q$_{\beta\beta}$ range (2039 keV) to check the results from the HdM experiment \cite{kla} within one year of data taking. Assuming a statistic of $\sim$15 kg$\cdot y$ and an energy resolution of 3.6 keV, the expected
number of background events is 0.5 counts. If no event is observed, a T$_{1/2}>3.0\cdot 10^{25}$ y (90\% C.L.) can be established with a detection efficiency of  95\%. This results in an upper limit on the effective neutrino mass
of m$_{ee}<0.2-0.8$ eV, depending on the nuclear matrix elements used. In GERDA Phase II, new diodes will be added to achieve 100 kg$\cdot$y of statistics within three years and new methods to reduce the background index by one order of magnitude to 10$^{-3}$ cts/(keV$\cdot$·kg$\cdot$·y) will be used. A third phase with $\sim$500 kg of target mass and a background index of 10$^{-4}$ cts/(keV$\cdot$·kg$\cdot$·y) is considered in the framework of a new worldwide collaboration. 

In 2005, the HdM and IGEX detectors were transported to the GDL at LNGS. GDL is a clean room of level 10 000 equipped with a clean bench and a radon-reduced bench of level 10. The performance parameters of the enriched detectors have been measured several times. In 2006, the diodes were taken out from their cryostat and sent to Canberra Semiconductor NV \cite{canberra} in preparation for GERDA Phase I. Before submerging the enriched diodes in liquid argon or nitrogen, a bare p-type HPGe diode has been used for tests in the GDL and at the detector manufacturer site. The prototype detector is mounted in the Phase I low-mass support and operated in liquid argon or liquid nitrogen. In total, 43 cooling/warming cycles have been carried out. Several modifications were done to the detector assembly and to the detector test stand to improve the detector performance. A long term measurement (2 months) in liquid argon was performed.

\section{Phase I enriched diodes}
To minimize the internal background of the detectors due to cosmogenic $^{68}$Ge and $^{60}$Co production, the germanium crystals must be stored underground. The IGEX detectors (RG 1-3) were transported from Canfranc Underground Laboratory, Spain, to LNGS in November 2005. For a long time, the IGEX detectors were kept warm with their cryostat out of liquid nitrogen. Therefore, it was necessary to perform a vacuum restoration cycle with each cryostat before operating the detectors. The cryostats were simultaneously heated and pumped. Afterward, the detectors were cooled down. As the high-voltage was applied, the resolution, the counting characteristics and the leakage current were measured. The performance parameters of the detectors were restored to their original values. If the diodes are stored for a long period of time at room temperature, an increase of the lithium layer thickness is expected, reducing the active mass of the detector. Several comparison measurements using different radioactive sources were performed to determine the thickness of the lithium dead layer of the IGEX detectors. Since their operation in IGEX experiment, no considerable changes of the active masses were observed.

The HdM detectors (ANG 1-5) stayed underground at LNGS since the start of the experi-ment. Except ANG 1, the detectors were stored at liquid nitrogen temperature. In 2005, their performance parameters (leakage current, operational voltage and energy resolution) were measured in GDL. A warming throughout pumping cycle was performed with ANG 1 cryostat to restore the original leakage current of the detector. All the HdM detectors were in working conditions, showing the same performance as $\sim$15 years ago.

\begin{figure}[h]
		\begin{centering}
		\includegraphics[scale=0.35]{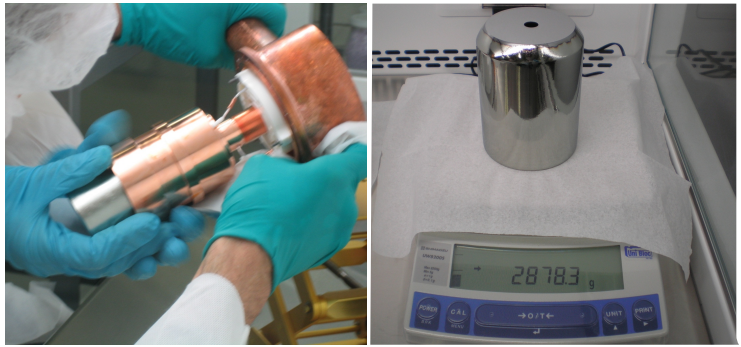}
	\caption{In 2006, the opening of the cryostats and the dimension measurements of the enriched diodes have been performed in the clean room environment of the GDL. Left: ANG 1 diode being taken out from its cryostat; right: measurement of the total mass of ANG 2 Ortec type diode.}
	\end{centering}
	\label{fig:dismounted}
\end{figure}

\begin{figure}[h]
		\begin{centering}
		\includegraphics[scale=0.3]{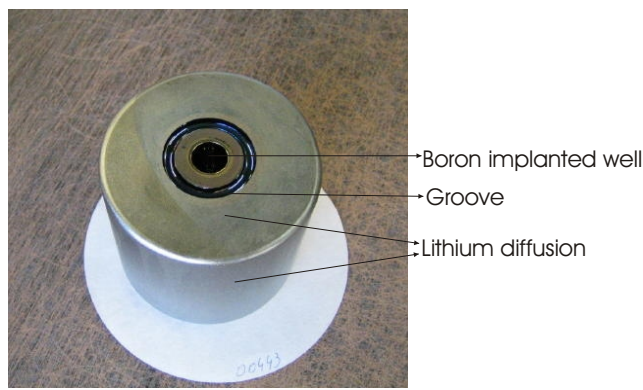}
	\caption{Canberra p-type germanium diode with a groove, a passivation layer covering the groove and a lithium diffused layer up to the groove.}
	\end{centering}
	\label{fig:cantype}
\end{figure}

In the year 2006, the cryostats have been opened and the enriched diodes dismounted in the clean room environment of the GDL. The total mass as well as the dimensions of the diodes and the cryostats have been measured (Figure \ref{fig:dismounted}). 
Table \ref{tab:hdm} gives the energy resolution of the HdM and IGEX detectors and the total mass of the diodes measured in the GDL. 
\begin{table}[h]
\begin{center}
\begin{tabular}{|| c | c | c | c | c | c | c | c |c ||} \hline\hline
 & ANG 1 & ANG 2 & ANG 3 & ANG 4 & ANG 5 & RG 1 & RG 2 & RG 3\\ \hline
FWHM at 1.3 MeV (keV) &2.9& 2.5& 3.0	& 2.8& 3.0 & 2.2& 2.3& 2.3\\
Mass (kg)& 0.97 & 2.88 & 2.45 &2.40& 2.78 & 2.15 & 2.19	& 2.12 \\  \hline\hline
       \end{tabular} 
        \end{center}
        \caption{The energy resolutions of the 5 Heidelberg-Moscow (measured in May 2005) and the 3 IGEX (measured in December 2005) detectors obtained in the GDL, together with the total mass of the diodes measured in 2006.}
        \label{tab:hdm}
\end{table}
After the dismounting, the diodes were put under vacuum in electro-polished stainless steel containers and transported to the detector manufacturer for refurbishment. Except ANG 1 which was a p-type Canberra diode, all the HdM and the IGEX enriched diodes were p-type Ortec diodes. The refurbishment consists of transforming all the enriched diodes into Canberra type diodes. Figure \ref{fig:cantype} shows a Canberra type germanium diode. 
\clearpage
Canberra type diodes have a groove at the p$^+$-n$^+$ junction which permits a p$^+$ contact outside the boron implanted well. The refurbishment process includes machining the groove, implanting the boron, diffusing the lithium layer up to the groove and evaporating the passivation layer. For most of the diodes, the refurbishment process is still ongoing. In-between the refurbishment steps, the diodes are stored underground at the HADES facility (500 mwe), Geel, Belgium. The total exposure to cosmic rays at surface level for the whole reprocessing and transportation process is $\sim$60 hours so the cosmogenic $^{60}$Co and $^{68}$Ge production is negligible. ANG 1 and RG 3 diodes have been refurbished. They are ready to be tested in the GDL test stand, once the series of measurements performed with the Phase I prototype detector is finished.

\section{Phase I prototype detector}
A non-enriched HPGe p-type diode (1.6 kg), which has been refurbished by Canberra Semiconductor NV using the same technology as for the enriched diodes, is used to investigate the performance of the Phase I detector assembly. The Phase I low-mass support, the cooling/warming cycles, the GDL liquid argon detector test stand and the detector stability in liquid argon have been tested.

The prototype detector is mounted in a low-mass holder (80 g) made of low-activity copper, Teflon and silicon. Only ultra-pure screened materials are used. The contribution of the detector holder to the background index is calculated using the screening limits in the Monte Carlo simulations ($<1.5\cdot 10^{-3}$cts/(keV$\cdot$kg$\cdot$y)). The detector assembly has first been tested at the detector manufacturer site. The mounting procedure, the signal and central high-voltage contact quality, the mechanical stability and the spectroscopy performance have been investigated. The same energy resolution as obtained in a standard test cryostat (2.2 keV at 1.332 MeV) has been obtained.

Since April 2006, the prototype detector is operated in the test stand of the GDL, consisting of an electro-polished dewar and high-purity liquid argon refilling system. The mounting of the detector assembly is performed inside the radon-reduced bench which is connected to the liquid argon detector test stand. The shielding of the test stand is made of 2.5 cm lead and 20 cm argon, suppressing the external background by a factor 10. The liquid argon level of the dewar is monitored by weighing cells. The radon concentration of the laboratory ($\sim$8 Bq/m$^3$) is monitored with a 3 l Lucas cell. The humidity of the laboratory is kept low (30\%).  For the first series of measurement, the prototype detector was operated in liquid nitrogen and since July 2006, the detector is operated in liquid argon. 

\begin{figure}[h]
		\begin{centering}
		\includegraphics[scale=0.2]{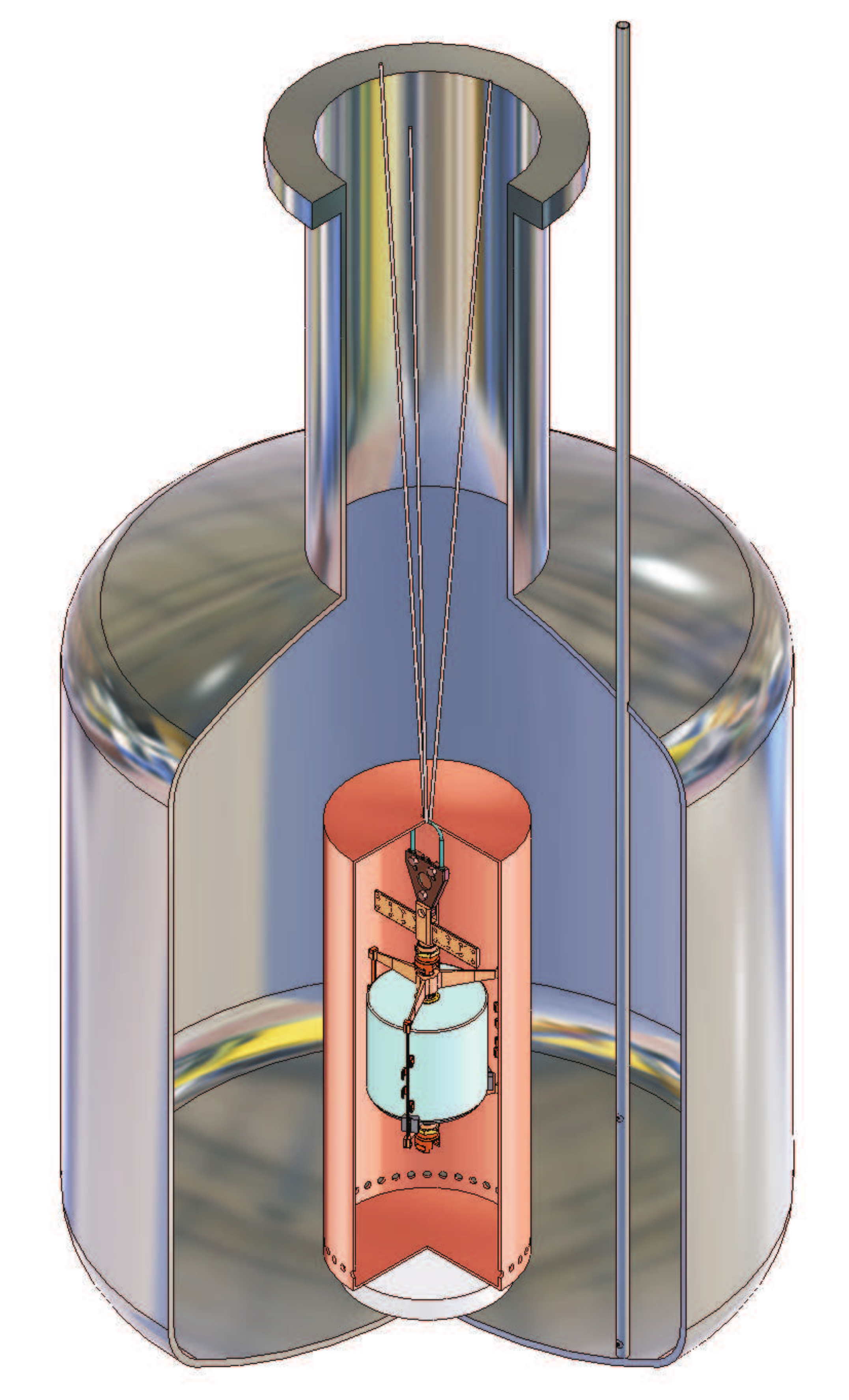}
	\caption{Liquid argon detector test stand of the GDL in which the GERDA Phase I prototype detector is operated.}
	\end{centering}
	\label{fig:dewar}
\end{figure}

Figure \ref{fig:dewar} shows a drawing of the 70 l dewar of the GDL test stand in which the Phase I prototype detector assembly is operated. It is equipped with a copper cylinder used as an infrared shield. In order to have the infrared shield always submerged, the liquid argon is topped-up every 4-5 days. The infrared radiation comes mainly from the neck of the dewar. In the present setup, the detector assembly is close enough to the dewar neck so it is sensitive to the infrared light. On the contrary, the detector array in GERDA will be sufficiently far away from the neck so no infrared shield will be needed.
The detector assembly is attached to the dewar flange on which a Canberra 2002 warm-FET preamplifier is installed. The signal cable is one meter long. The leakage current is monitored with the test point voltage and/or with a picoampermeter. A standard ORTEC amplifier 672 and ADC 919 are used to collect the spectra. The energy resolution obtained with this setup is $\sim$3 keV FWHM at 1.332 MeV.

Several measurements are performed to monitor the conditions of the detector assembly. Before mounting the diode in its low-mass support, the diode is cooled down in a special holder and the leakage current is measured as a function of the applied high-voltage. The diode is warmed up, mounted in its support and cooled down in the test stand dewar. The quality of the signal and high-voltage contacts is monitored before, during and after the cooling process. Subsequently, the leakage current and the energy resolution are recorded as the high-voltage is applied to the detector.

Within one year of testing, the Phase I prototype detector assembly has been operated in liquid argon and nitrogen, 43 cooling-warming cycles have been performed to do mounting and/or electronics modifications. Twice, the prototype detector showed high leakage current so it was transported to the detector manufacturer to evaporate a new passivation layer. A long term measurement (2 months without doing any modifications to the setup) has been performed and the detector parameters were stable. 

Beyond the original purpose of the long term stability measurement, 10 days of background measurement were used to estimate the sensitivity of the present setup to the neutrinoless double electron capture (0$\nu$ECEC) process of $^{36}$Ar. Natural
argon contains the isotope $^{36}$Ar with an abundance of 0.336\%, which is expected to be unstable undergoing double electron capture (ECEC) \cite{ecec}. No measurements of the half life limit
exist. In the 0$\nu$ECEC process, two orbital
electrons are absorbed by the nucleus. The momentum-energy
conservation requires the released energy Q to be emitted with some additional particle(s). We consider the neutrinoless process in which the released energy is carried away by
photons. The experimental signature of this decay consists of the emission of three photons
with total energy Q (433.5 keV): two X-rays with energies of the corresponding holes in the electron shells
of the daughter atom produced by the ECEC capture, and one photon taking the rest of the
available energy. This photon can be detected by a
high resolution germanium detector. The radiative 0$\nu$ECEC process signature is a sharp peak in the area of the Q value of the ECEC reaction. The limit obtained for the 0$\nu$ECEC process of $^{36}$Ar with the emission of a single photon
is T$_{1/2}(0^+{\rightarrow}$g.s. with three photons) $\geq 1.9\cdot10^{18}$ years (68\% C.L.). The sensitivity of the experiment presented here is limited by external radiation of the
detector test stand which is not designed as a low-background setup.

\section{Conclusion}
The GERDA experiment will be located at LNGS underground laboratory to search for 0$\nu\beta\beta$ decay of $^{76}$Ge. Bare
HPGe detectors enriched in $^{76}$Ge will be submerged in liquid argon serving
simultaneously as a shield against external radioactivity and a cooling medium. GERDA will start with the enriched diodes from the previous IGEX and HdM experiments. After one year of data taking, GERDA aims to check the claim on the 0$\nu\beta\beta$ decay evidence of HdM $^{76}$Ge experiment with an exposure of $\geq$15 kg$\cdot$y and a background index of 10$^{-2}$ cts/(keV$\cdot$·kg$\cdot$·y). In 2005, the performance of the HdM and IGEX detectors has been measured in the GDL. After $\sim$15 years of operation and storage, the detector performances were unchanged. In the preparation for GERDA Phase I, the IGEX and HdM cryostats have been opened, the enriched diodes dismounted and sent to Canberra Semiconductor NV for refurbishment. Since January 2006, tests with a bare HPGe prototype detector in the GDL and at the detector manufacturer site are being carried out. The low-mass holder, the cooling/warming cycles, the stability in liquid argon and the refurbishment procedure have been investigated. A long term measurement was performed and the detector parameters were stable. Additionally, the first limit on the 0$\nu$ECEC process of $^{36}$Ar was derived.

\section{Acknowledgments}
This work was partly supported by INTAS (Project Nr. 05-1000008-7996). M. Barnab\'e Heider acknowledge the financial support from NSERC and A. Smolnikov and S. Vasiliev acknowledge the support from RFBR (grant 06-02-01050).

\end{document}